\documentstyle[12pt]{article}

\newcommand{\be}{\begin{equation}}
\newcommand{\ee}{\end{equation}}
\newcommand{\bea}{\begin{eqnarray}}
\newcommand{\eea}{\end{eqnarray}}

\newcommand{\tZ}{\tilde Z}

\newcommand{\tL}{\tilde L}

\newcommand{\w}{\wedge}
\newcommand{\ch}{{\rm ch}}
\newcommand{\tr}{\triangle}

\makeatletter
\newdimen\normalarrayskip              
\newdimen\minarrayskip                 
\normalarrayskip\baselineskip
\minarrayskip\jot
\newif\ifold             \oldtrue            

\makeatother

\newlength{\extraspace}
\newlength{\extraspaces}
\begin{document}

\addtolength{\baselineskip}{.8mm}

\thispagestyle{empty}

\begin{flushright}
\baselineskip=12pt
{\sc OUTP}-98-32P\\
hep-th/9805027\\
\hfill{  }\\September 1998
\end{flushright}
\vspace{.5cm}

\begin{center}

\baselineskip=24pt

{\Large\bf{Black String Entropy \\from Anomalous D-brane Couplings
}}\\[15mm]

\baselineskip=12pt

{\sc David D. Song\footnote{E-mail: {\tt d.song1@physics.oxford.ac.uk}}}
{\sc and Richard J.\ Szabo\footnote{Work supported in part by PPARC
(U.K.); Address after September 1, 1998: The Niels Bohr Institute,
University of Copenhagen, 
Blegdamsvej 17, DK-2100 Copenhagen \O, Denmark \\
E-mail: {\tt szabo@alf.nbi.dk}}}
\\[5mm]

{\it Department of Physics\\ University of Oxford\\ Oxford OX1 3PU, U.K.}
\\[15mm]

\vskip 1.5 in

{\sc Abstract}

\begin{center}
\begin{minipage}{15cm}

The quantum corrections to the counting of statistical entropy for the
5+1-dimensional extremal black string in type-IIB supergravity with two 
observers are studied using anomalous Wess-Zumino actions for the
corresponding intersecting D-brane description. The electric-magnetic
duality symmetry of the anomalous theory implies a new symmetry between
D-string and D-fivebrane sources and renders opposite sign for the RR
charge of one of the intersecting D-branes relative to that of the black
string. The electric-magnetic symmetric Hilbert space decomposes into
subspaces associated with interior and exterior regions and it is shown
that, for an outside observer, the expectation value of a horizon area
operator agrees with the deviation of the classical horizon area in going
from extremal to near-extremal black strings.
\end{minipage}
\end{center}

\end{center}

\noindent
\vfill
\newpage
\pagestyle{plain}
\setcounter{page}{1}

A startling consequence of the D-brane description of solitonic states
\cite{polchinski} is the ability to microscopically count the statistical
entropy of black holes in string theory. By realizing that intersecting
D-branes are the weak coupling limit of certain five-dimensional
supergravity black hole solutions and carry the same Ramond-Ramond (RR)
charge as that of the black hole, the counting of quantum states from
BPS-saturated D-brane bound
states can be shown to agree with the Bekenstein-Hawking formula
(proportional
to the horizon area) for the entropy \cite{vafa,hor} (see \cite{mal} for a
review). In this letter we shall study quantum corrections to a class of 
classical black hole solutions in the D-brane description by considering
the 
points of view of two observers where one is the observer outside the
horizon 
and the other is a ``superobserver'' who, as advocated in the proposal of
black hole complementarity \cite{susskind}, sees both exterior and
interior regions 
of black holes ({\it i.e.} the Hilbert space of both exterior and interior 
regions is relevant).  A natural way to understand the
black hole/D-brane correspondence is to identify a quantum area
operator on D-brane Hilbert space whose classical limit yields the horizon
area
of the extremal black hole in $N=8$ supergravity \cite{mar}. The
five-dimensional black hole can be viewed as a six-dimensional black
string
which winds around a compact internal circle. The horizon area of the
extremal
black string and the expectation value of the area operator agree only in
the
limit of large winding number \cite{mar}. In an effort to find the
location of the quantum states responsible for entropy counting,
it was shown in \cite{mar2} that the D-strings and D-fivebranes carrying
the
same RR charges as the black strings can be placed at the singularity of
the interior region of the black string solution and argued that their
modes stretch from the singularity out to the horizon.

When D-branes intersect on a topologically non-trivial manifold, the
twistings
of their normal bundles yield chiral anomalies in their worldvolume field
theories which induce anomalous couplings of the D-brane gauge and
gravitational fields to the RR tensor potentials
\cite{moore}--\cite{moore2}.
The induced Wess-Zumino terms imply that topological defects (such as
instantons or monopoles) on the D-branes carry their own RR charge
determined
by their topological quantum numbers \cite{witten}. In the following we
will
show that, when one carefully takes into account of the anomalies which
occur
on intersecting D-branes, while the RR charge of one D-brane coincides
with the RR charge of black string, the RR charge of the other
intersecting D-brane is of opposite sign.

Following \cite{mar2}, we interpret this effect as placing, at strong
coupling, one D-brane in the interior and the other in the exterior region
of the black string horizon. This
change arises from the fact that the anomalous Wess-Zumino terms should be
properly understood \cite{yin} as part of a total action which treats
electric
and magnetic potentials and sources on equal footing. This
electric-magnetic
symmetry decomposes the total Hilbert space symmetrically in terms of
the individual D-brane Hilbert spaces. This leads to an intriguing
symmetry 
between interchange of the two intersecting D-branes and hence of the
interior 
and exterior regions of the black hole.\footnote{This symmetry is not to
be 
confused with the S-duality symmetry of type-IIB superstring theory.}  We
shall argue that this new interpretation leads to 
the explanation of the discrepancy raised in the black hole/D-brane 
correspondence \cite{mar}.  Indeed, this new picture leads us to consider
both 
superobserver's and outside observer's points of view and we will show
that, 
for an outsider observer, this interpretation effectively perturbs the 
classical extremal black string solution to the near-extremal case. The
horizon
area from this modified black string solution agrees with the expectation
value of the area operator on D-brane Hilbert space. Thus the proper
incorporation of anomalous D-brane couplings sheds new light on the role
of
observers in the quantum D-brane picture.

We start by briefly reviewing some properties of six-dimensional black
strings
in type IIB string theory. The bosonic action for type IIB supergravity in
the
Einstein frame is of the form
\be
\frac{1}{16 \pi G_{10}} \int_{{\cal M}_{10}} d^{10}x
{}~\sqrt{-g}\left(R-\frac{1}{2}(\nabla
\phi)^{2} - \frac{1}{12} e^\phi H_{(3)}^2\right)+\dots
\label{IIBaction}\ee
where the dots denote non-covariant field terms, $H_{(3)}$ is the RR
three-form
field strength, $\phi$ is the dilaton field, and $G_{10}$ is the
ten-dimensional gravitational constant which in string units is given by
$G_{10}=8\pi^6g_s^2$ with $g_s$ the string coupling constant. The black
string
solution is most easily constructed on the ten-dimensional spacetime
${{\cal
M}_{10}}=T^5\times R^5$ with the five-torus $T^5$ taken to be the product
of a
circle $S^1$ (along which the string lies), with coordinate $z$ and
circumference $L$, and a four-torus $T^4$ with coordinates $y^i$ and
hypervolume $V$. The Cartesian coordinates of the remaining 4+1
asymptotically
flat non-compact directions are denoted by $(x^i,t)$. In the limit $L \gg
V^{1/4}$ the extremal black string solution in 5+1 dimensions reduces to
the
4+1 dimensional Reissner-Nordstr\"om black hole of type IIB superstring
theory.

In light-cone coordinates $u \equiv t-z, v \equiv t+z$, the classical
six-dimensional exterior and interior black string solutions of the
coupled
Einstein equations which follow from the action (\ref{IIBaction}) are
\cite{mar2}
\bea
ds_{\rm ext}^2 &=& h_1^{1/4}h_5^{3/4}\left[\frac{du}{h_1h_5}(-dv+Kdu) +
\frac{dy_idy^i}{h_5}+dx_idx^i\right]
\label{extmetric}\\ds_{\rm
int}^2&=&h_1^{1/4}h_5^{3/4}\left[\frac{du}{h_1h_5}(dv+Kdu)+\frac{dy_idy^i}{h_5}
+dx_idx^i\right]\label{intmetric}\\e^{-2\phi}&=&\frac{h_5}{h_1}\label{dilaton}
\\H_{(3)\nu uv}&=&h_1^{-2}\partial_\nu h_1 \ \ \ , \ \ \
H_{(3)ijk}~=~-\epsilon_{ijkl}\partial_{x_l}h_5
\label{3form}\eea
where the harmonic functions are
\be
h_1 = 1+\frac{r_1^2}{r^2} \  \ , \  \  h_5=1+\frac{r_5^2}{r^2}
\label{harmfnsext}\ee
for the exterior metric (\ref{extmetric}) and
\be
h_1 =-1+\frac{r_1^2}{r^2} \  \ , \  \ h_5=-1+\frac{r_5^2}{r^2}
\label{harmfnsint}\ee
for the interior metric (\ref{intmetric}), with $r^2=x_ix^i$. The horizon
lies 
at the coordinate singularity $r=0$ in (\ref{extmetric}, \ref{intmetric})
and 
the timelike curvature singularity lies at either $r=r_1$ (if $r_1<r_5)$
or 
$r=r_5$ (if $r_5<r_1)$.  The constants $r_1$ and $r_5$ determine the
electric 
and magnetic charges of the black hole with respect to the RR three-form
field 
strength by
\bea
Q_1 & \equiv &
\frac{V}{(2\pi)^6
g_s}\int_{S^3}e^{\phi}*_6H_{(3)}~=~\frac{Vr_1^2}{(2\pi)^4 
g_s}
\label{Q1}\\
Q_5 & \equiv & \frac{1}{(2\pi)^2 g_s} \int_{S^3} H_{(3)} ~ = ~ 
\frac{r^2_5}{g_s}
\label{Q5}\eea
where $S^3$ is a three-sphere and $*_6$  the Hodge dual in the
six-dimensional spacetime $R^5\times S^1$. With this normalization the
charges
$Q_1$ and $Q_5$ are integers. For simplicity, we will consider only the
homogeneous case where $K=p/r^2$ with $p$ the momentum of the black hole.
Note that, aside from the
forms of the harmonic functions, the only difference between the exterior
and
interior solutions is the change in sign of the $dudv$ term in the metric.

By putting a $u$ dependence on the momentum $p\equiv p(u)$, one can add 
travelling waves \cite{mar3}, {\it i.e.} waves moving along
the string, since $p(u)$ then changes the momentum density along the
string.
For
extremal black holes, there are only left-moving modes, and therefore $p$
depends on $u$ but not on $v$. The corresponding horizon area is
\be
 A= 2\pi^2 r_1 r_5 V \int \sqrt{p(u)} ~du
\label{area}\ee
When $p \equiv (2\pi)^6 g_s^2N/L^2V$ is constant, with $N$ the number of
modes,
(\ref{area}) yields the familiar extremal Bekenstein-Hawking entropy
\be
S_e=\frac{A}{4G_{10}}=2\pi\sqrt{Q_1Q_5 N}
\label{extentropy}\ee
of the 4+1 dimensional black hole.

Being sources which carry RR charge, the electric and
magnetic charges can be induced by superposing $Q_1$ D-string and $Q_5$
D-fivebrane sources with worldvolumes $M_1$ and $M_5$ lying in the $(t,z)$
plane $R^1\times S^1$ and the 5+1 dimensional $(t,z,y^i)$ space $R^1\times
S^1\times T^4$, respectively. The D-strings and D-fivebranes intersect
along
the worldsheet of the string. Since the D-strings wind $Q_1$ times around
the
$S^1$ and the D-fivebranes wind $Q_5$ times around the whole $T^5$, the
effective string wraps $Q_1Q_5$ times around the compact $z$-direction.
This
feature is the key to counting the entropy (\ref{extentropy}) using
D-branes.  
Using the positive energy theorem for black
holes, it is argued in \cite{mar2} that outside the horizon, BPS D-brane
solutions with positive charge and energy density are static whereas
D-branes
with negative charge and energy density are unphysical. Inside the
horizon, the
situation is opposite in that it is the negatively charged D-brane or
anti-D-brane which is static (and has positive energy density).

The comparison of black hole states to quantum configurations of the
D-branes
can be made by constructing an area operator for the black string on the
D-brane Hilbert space ${\cal H}_D$ \cite{mar}. This is achieved by
defining a
momentum operator and plugging it into (\ref{area}). We will consider the
case
when $\sqrt{p} \ll r_1, r_5$.  The ensemble of
homogeneous black string states contains those states which lie on a
single
string of tension $1/2\pi g_sQ_5$ and within a length scale $\tilde
L\equiv
Q_1Q_5L$. The string can be described by a free worldsheet $\sigma$-model
for a
single bosonic field $X$, which in the limit $L\gg V^{1/4}$ describes the
excitations of the D-branes. Then the momentum density along the string
can be
written as $T_{++}=(\partial_+X)^2/2\pi g_sQ_5$, whose mode expansion is
given
by $\partial_+X=(\sqrt{2\pi^2g_sQ_5}/\tilde{L})\sum_{m=-\infty}^{\infty}
\alpha_me^{-2\pi im\sigma^+/\tilde{L}}$ with
$[\alpha_m,\alpha_n]=m\delta_{-m,n}$. Since the differences between the
worldsheet coordinate $\sigma^+$ and the spacetime coordinate $u$ are
integer
multiples of $L$ due to the string wrapping around the compact direction
$Q_1Q_5$ times, it follows that the parameters of the black string
solution can
be identified with quantum fields on the effective string worldsheet
through
the momentum operator
\be
p(u) = \frac{1}{\tL} \sum^{\infty}_{l=-\infty} p_l e^{-2\pi i lu/\tL}
\label{momop}\ee
where
\be
p_l = \frac{2\pi^2g_s^2}{V\tL}\sum^{Q_1Q_5}_{k=1}
\sum^{\infty}_{n=-\infty}
e^{-2\pi i k l L/\tL}:\alpha_{l-n} \alpha_n:
\label{pl}\ee
Note that in (\ref{pl}) the index $k$ runs from 1 to $Q_1Q_5$.

Using the equipartition theorem, it can be shown \cite{mar} that the
difference
between the expectation value of the area operator (\ref{area}) in ${\cal
H}_D$
and the area of the classical stationary black string can be estimated by
evaluating the quantity $\sum_{k\neq 0}\frac{<:p_kp_{-k}:>}{<:p^2_0:>}$.
The
quantum fluctuations in the longitudinal momentum yield the deviation
\be
\sum_k \frac{<:p_kp_{-k}:>}{<:p^2_0:>} \sim \frac{1}{Q_1 Q_5}
\label{dev}\ee
from the area of the stationary black hole \cite{mar}. Therefore for large
winding number $Q_1Q_5$, the two areas can match. But generally there is
always
a deviation between the classical horizon area and the expectation value
of the
area operator on ${\cal H}_D$ which leads to an entropy counting
difference
\be
\frac{\Delta S}{S_e}\sim\frac1{\sqrt{Q_1Q_5}}
\label{entropydev}\ee
between the classical and (quantum) D-brane approaches.

In the following we will describe an origin for the discrepancy
(\ref{entropydev}). D-brane field theory plays a vital role in the
counting of
statistical entropy for a quantum description of black holes at strong
coupling
\cite{vafa}--\cite{mal}. At low energy, these field theories have both
gauge
and global symmetries. Anomalies arise due to the chiral asymmetry, with
respect to the global R symmetry, of massless Weyl fermion fields on the
intersection of D-branes. These anomalies can be compensated for by the
anomalous variation of the classical D-brane actions. In
\cite{moore}--\cite{moore2} it was shown that the classical variations of
Wess-Zumino actions for the D-branes cancel the Yang-Mills and
gravitational
anomalies as well as the anomalies associated with global R symmetries.
Let us
now quickly describe how this works.

The low-energy dynamics of the configuration of $Q_1$ D-strings and $Q_5$
D-fivebranes, each with infinitesimal separation, are described,
respectively,
by supersymmetric field theories on the worldvolumes $M_1$ and $M_5$ which
have
$U(Q_1)$ and $U(Q_5)$ gauge symmetries \cite{polchinski,wittenD}. The
anomalous
gauge variation on the intersection of D-strings and D-fivebranes on the
effective string worldsheet $M_1\cap M_5$ can be written as
\cite{moore}--\cite{moore2}
\be
2\pi\int_{{\cal M}_{10}}\tr_{M_1} \w \tr_{M_5} \w \left( \ch(F_1) \w
\ch(-F_5) \w\frac{{\hat A}[T(M_1) \cap T(M_5)]}{{\hat A}[N(M_1) \cap
N(M_5)]}\right)^{(1)}
\label{anombl}\ee
where the delta-function supported Poincar\'e dual form $\tr_{M_i}$, of
degree
$10-\dim M_i$, is the de Rham current of $M_i$, i.e.
$\int_{M_i}Z\equiv\int_{{\cal M}_{10}}\tr_{M_i}\w Z$. If $N$ is the
leading
constant part of the closed gauge-invariant form $Z\equiv N+dZ^{(0)}$,
then
$Z^{(1)}$ denotes its corresponding Wess-Zumino descendent defined by the
first-order gauge variation $\delta_gZ^{(0)}\equiv dZ^{(1)}$ of its
secondary
characteristic. The tangent bundle $T({{\cal M}_{10}})$ of the total space
decomposes locally as $T(M_i)\oplus N(M_i)$, with $N(M_i)$ the normal
bundle of
the embedded submanifold $M_i$, and ${\hat A}$ denotes the Dirac genus.
$F_1$
and $F_5$ are Hermitian $U(Q_1)$ and $U(Q_5)$ gauge field strengths,
respectively, and
\be
\ch(F_j) \equiv {\rm tr}_{Q_j}
\exp\frac{iF_j}{2\pi}=Q_j+\ch_1(F_j)+\ch_2(F_j)+\dots
\label{ch}\ee
are the corresponding Chern characteristic classes in the fundamental
$Q_j$
representations of the respective gauge groups on the D-branes. The
anomaly
term (\ref{anombl}) comes from the tensor product of the spinor bundles,
associated with the lifting of $T({{\cal M}_{10}})=T(M_i)\oplus N(M_i)$,
with
the Chan-Paton vector bundles over the D-branes. As shown in \cite{moore},
the
anomalous zero modes on the intersection of the D-branes come from the
massless
excitation spectrum of the D-brane field theory which consists of Weyl
fermions
in the mixed sector $Q_1\otimes\bar Q_5$ and $\bar Q_1\otimes Q_5$
representations of the gauge group $U(Q_1)\times U(Q_5)$. Since the Chern
characteristic class for the adjoint representation of the unitary group
can be
decomposed as $\ch[U(Q_i)_{Q_i\otimes \bar
Q_i}]=\ch(F_i)\w\ch(F_i^*)=\ch(F_i)\w\ch(-F_i)$, the $\ch(F_1)\w\ch(-F_5)$
term
comes from the mixed sectors $Q_1\otimes{\bar Q}_5$ and ${\bar Q}_1\otimes
Q_5$
which both contribute equally.

The anomaly term (\ref{anombl}) can be cancelled by carefully writing the
constituent bulk D-brane actions in well-defined forms. The Wess-Zumino
part of
the low-energy effective action for the D-brane dynamics has the form
\be
I_{\rm WZ}=-\frac{\mu}{2}\sum_{i=1,5}\int_{M_i} C \w Z_i
\label{cs}\ee
where $\mu$ is the D-brane charge, $C$ is the sum of all the even
form-degree
RR gauge fields for the type-IIB theory we are considering here, and $Z_i$
are
the D-brane sources which are invariant polynomials of the Yang-Mills
field
strengths and gravitational curvatures on $M_i$. Denoting the $p$-form
parts of
$C$ and $Z_i$ by $C_{(p)}$ and $Z_{i(p)}$, the action (\ref{cs}) can be
written
as an integral over the total spacetime using the de Rham currents as
\bea
I_{\rm WZ}&=&-\frac{\mu}{2}\int_{{\cal
M}_{10}}\left\{\tr_{M_1}\w\left(C_{(0)}\w Z_{1(2)}+C_{(2)}\w
Z_{1(0)}\right)\right.\label{csdeRham}\\&
&~~~~~~~~~~+\left.\tr_{M_5}\w\left(C_{(0)}\w Z_{5(6)}+C_{(2)}\w
Z_{5(4)}+C_{(4)}\w Z_{5(2)}+C_{(6)}\w Z_{5(0)}\right)\right\}\nonumber
\eea
where we recall that $\tr_{M_1}$ is an eight-form and $\tr_{M_5}$ is a
four-form.

However, because of the D-brane couplings, the gauge fields are not
globally-defined as single-valued differential forms, and (\ref{cs})
should be
written more carefully since $H$, the total RR field strength, has global
corrections to its local form $dC$. The appropriate modification is the
anomalous D-brane coupling
\be
I_{\rm WZ}=-\frac{\mu}{2}\sum_{i=1,5}\int_{{\cal
M}_{10}}\tr_{M_i}\w\left(N_iC-H\w Z_i^{(0)}\right)
\label{modcs}\ee
which coincides with (\ref{cs}) upon integration by parts when $H=dC$, but
otherwise contains global corrections to the form (\ref{cs}). The
equations of
motion which follow from coupling the kinetic term of (\ref{IIBaction}) to
the 
source action (\ref{modcs}) are
\be
d(e^{\phi}*_{10}H) =\mu\sum_{i=1,5}\tr_{M_i}\w Z_i
\label{eqnmot}\ee
where $*_{10}$ is the Hodge dual in the ten-dimensional spacetime ${\cal
M}_{10}$, and the Bianchi identities are \cite{yin}
\be
dH=-\mu\sum_{i=1,5}\tr_{M_i}\w\tZ_i
\label{bianchi}\ee
where $\tilde Z_{i(l)}=(-1)^{(\dim M_i+l+2)/2}Z_{i(l)}$ is the conjugate
to
$Z_i$ defined by reversing the orientations of the Chan-Paton and tangent
bundles over the D-branes. The gauge variation of the potential is thus
$\delta_gC=\mu\sum_i\tr_{M_i}\w\tZ_i^{(1)}$, and it follows that the gauge
variation of the modified Wess-Zumino action (\ref{modcs}) is
\be
\delta_gI_{\rm WZ}=-\frac{\mu^2}{2}\int_{{\cal M}_{10}}\tr_{M_1} \w
\tr_{M_5}
\w\left(Z_1 \w \tZ_5 +Z_5 \w \tZ_1\right)^{(1)}
\label{gaucs}\ee
Therefore with
\bea
Z_i&=&\ch(F_i)\w\sqrt{\frac{\hat{A}[T(M_i)]}{\hat{A}[N(M_i)]}} \label{ZZ}
\\
\tZ_j&=&\ch(-F_j) \w \sqrt{\frac{\hat{A}[T(M_j)]}{\hat{A}[N(M_j)]}}
\label{ZZt}\eea
and $\mu^2=2\pi$, the variation (\ref{gaucs}) cancels the anomaly term
(\ref{anombl}).

The Bianchi identity (\ref{bianchi}) follows from the fact that the
anomalous 
coupling (\ref{modcs}) explicitly involves both electric and magnetic
sources 
and is to
be understood as part of an action which is manifestly electric-magnetic
symmetric \cite{yin}. As we will now demonstrate, the anomalous action
(\ref{modcs}) with this underlying electric-magnetic duality yields new
implications for the quantum descriptions of black holes in the D-brane
picture. We note first of all that the field strength $H$ in
(\ref{eqnmot}, \ref{bianchi}) is the sum of all the odd form-degree RR
field
strengths for the type-IIB theory. To compare the physical D-brane charges
derived from the anomalous coupling (\ref{modcs}) with those of the black
string solution of (\ref{IIBaction}), we are interested in the three-form
component $H_{(3)}$ of $H$. We therefore consider two hypervolumes $B^8$
and
$B^4$ in ${\cal M}_{10}$ whose boundaries are spheres $S^7$ and $S^3$
enclosing
the total D-string and D-fivebrane charge. From (\ref{eqnmot})  and
(\ref{bianchi}) we have\footnote{Note that here, in contrast to the RR
charges
(\ref{Q1}) and (\ref{Q5}) which are defined in the six-dimensional black
string
frame, we equivalently define D-brane charge in a ten-dimensional frame.}
\bea
\int_{B^8} d(e^{\phi}*_{10} H_{(3)}) & = & \mu\int_{B^8}\left(\tr_{M_1}\w 
Z_{1(0)} + \tr_{M_5}\w Z_{5(4)}\right) \label{QD1} \\
\int_{B^4} dH_{(3)} & = & -\mu\int_{B^4}\tr_{M_5}\w \tZ_{5(0)} \label{QD5}
\eea
The first term on the right-hand side of (\ref{QD1}) corresponds to the
physical charge ${\cal Q}_1$ of the D-strings, while the second term is an
induced anomaly charge. This latter charge arises from the fact
\cite{witten}
that RR charge conservation requires the boundaries of the D-string to
carry
instanton number with respect to the D-fivebrane worldvolume gauge field.
The
right-hand side of (\ref{QD5}) represents the physical charge ${\cal Q}_5$
of
the D-fivebranes.

For the D-string and D-fivebrane worldvolumes we have $\tilde
Z_{i(0)}=Z_{i(0)}$. It follows from (\ref{ZZ}), (\ref{ch}) and the fact
that
the constant part of the Dirac genus is 1 that
\bea
{\cal Q}_1&=&\mu Q_1\label{QD1phys}\\{\cal Q}_5&=&-\mu Q_5\label{QD5phys}
\eea
We see that, for $\mu=+\sqrt{2\pi}$, the RR charge of D-strings coincides
with (\ref{Q1}) while the RR charge of D-fivebranes is of opposite sign to
(\ref{Q5}).\footnote{For $\mu=-\sqrt{2\pi}$, RR charge of D-strings would
have opposite sign to (\ref{Q1})}  This would seem to contradict that
D-branes, at weak coupling, carry the same RR charge as that of the black
hole.

The crucial minus sign which appears in (\ref{QD5phys}) comes from the
electric-magnetic duality property of the anomalous Wess-Zumino coupling
(\ref{modcs}). The total (non-covariant) explicitly duality symmetric
action
$I_{BE}+I_{\rm WZ}$ comes from writing the field strength term in the
supergravity action (\ref{IIBaction}) as \cite{yin,emduality}
\be
I_{BE}=-\frac1{384\pi G_{10}}\int_{{\cal M}_{10}}e^\phi\left(B_{(3)}\wedge
E_{(7)}+B_{(7)}\wedge
E_{(3)}+B_{(3)}\wedge*_{10}B_{(3)}+B_{(7)}\wedge*_{10}B_{(7)}\right)
\label{emdualaction}\ee
where we have decomposed $H_{(p)}=E_{(p)}+B_{(p)}$ for $p=3,7$ in terms of
its
components $E_{(p)}$ with a temporal index and its components $B_{(p)}$
involving only spatial indices on ${\cal M}_{10}$. Thus when one
interprets the 
D-string and D-fivebrane fields as ten-dimensional
electromagnetic duals of one another, the physical couplings of the black
string change. This symmetry is necessary to cancel the anomalous
fermionic 
zero modes (\ref{anombl}) via a change of orientation of the bundles over
one 
of the intersecting D-branes.

Generally, the construction of explicitly electric-magnetic symmetric
quantum
field theories involves augmenting the Hilbert space so as to include two
independent gauge potentials where one is the physical field and the other
is the unphysical dual field \cite{emduality,lizzi}. In the present case
the two
potentials are associated each with the D-string and D-fivebrane in ${\cal
M}_{10}$, and corresponding sources $Z_i$ and $\tilde Z_i$, which in turn
means
that the full D-brane Hilbert space to be considered is the tensor product
\be
{\cal H}_D={\cal H}_{D1}\otimes{\cal H}_{D5}
\label{hilbertsp}\ee
of independent D-string and D-fivebrane Hilbert spaces. Since the Hilbert
space of the exterior and interior regions of the black hole is relevant
only to a superobserver, one could associate the unphysical dual field
with the interior region of the black hole for an outside observer, i.e.
one of the intersecting D-branes is associated with exterior while the
other is associated with interior region of black string. It was argued
that \cite{mar2} in placing D-branes in static equilibrium on the extremal
black hole, outside the horizon the positively charged D-brane is static
whereas inside the horizon the negatively charged D-brane is static.
Therefore, if we consider that, at strong coupling and for
$\mu=+\sqrt{2\pi}$ (which we henceforth assume), the D-string is placed in
the exterior while the D-fivebrane should be interpreted as lying on the
interior of the black string horizon, then  this  resolves the sign
difference between the RR charges (\ref{QD5phys}) and (\ref{Q5}).

The decomposition of the Hilbert space (\ref{hilbertsp}) can then also be
thought of as yielding Hilbert
spaces ${\cal H}_{D1}$ for the exterior and ${\cal H}_{D5}$ for the
interior
regions of the quantum black hole. This augmented Hilbert space of
exterior and interior regions can be of use only to a superobserver
\cite{susskind}.  From a superobserver's point of view, both the D-string
and D-fivebrane are positively charged and only left moving momentum modes
exist. This augmented Hilbert space also gives a remarkable manifestation
of the electric-magnetic symmetry as an 
interior-exterior region symmetry of the black string. The horizon, where
the 
D-branes intersect, then corresponds to
projecting the Hilbert space (\ref{hilbertsp}) onto only one component
representing the physical (duality non-symmetric) solution in which both
D-branes are placed in the same region. With this projection onto
``physically 
observable'' states ({\it i.e.} those relevant to an outside observer),
the RR 
three-form field strength and its dual are related by
$H_{(3)}=-*_{10}H_{(7)}$ 
\cite{yin}, which gives $E_{(7)}=-*_{10}B_{(3)}$ and
$E_{(3)}=-*_{10}B_{(7)}$ in (\ref{emdualaction}) leaving only one set of
electromagnetic tensor fields for the physical black string solution.

This interpretation also modifies the momentum operator (\ref{momop}) on
the
D-brane Hilbert space, which with respect to the decomposition
(\ref{hilbertsp}) can be considered as a function $p(u_1,u_5)$ of two
variables
$u_1$ and $u_5$ representing the worldsheet coordinates of the D-string
and
D-fivebrane, respectively. The corresponding mode expansion 
(\ref{momop},\ref{pl}) can be rewritten as
\be
p(u_1,u_5)= \frac{1}{\tL}\sum_{l_1,l_5=-\infty}^\infty(p_{l_1}\otimes 
p_{l_5})e^{-2\pi
i(l_1u_1+l_5u_5)/\tL}
\label{pl15}\ee
where
\bea
p_{l_1}&=&\sqrt{\frac{2\pi^2g_s^2}{V\tL}}\sum^{Q_1}_{k_1=1}
\sum^{\infty}_{n_1=-\infty}  e^{-2 \pi ik_1l_1 L/\tL}
:\alpha_{l_1-n_1}^{(1)}\alpha_{n_1}^{(1)}: \nonumber \\
p_{l_5}&=&\sqrt{\frac{2\pi^2g_s^2}{V\tL}}\sum^{Q_5}_{k_5=1}
\sum^{\infty}_{n_5=-\infty}  e^{-2\pi ik_5l_5 L/\tL}:\alpha_{l_5-n_5}
^{(5)}\alpha_{n_5}^{(5)}:
\label{plmodes}\eea
and the modes $\alpha_{n_1}^{(1)}$ and $\alpha_{n_5}^{(5)}$ act on the
Hilbert
spaces ${\cal H}_{D1}$ and ${\cal H}_{D5}$, respectively. Note that the
index
$k_1$ runs from 1 to $Q_1$ for $p_{l_1}$ while $k_5$ runs from 1 to $Q_5$
for
$p_{l_5}$.  The modification (\ref{pl15},\ref{plmodes}) of 
(\ref{momop},\ref{pl}) does not change the deviation (\ref{dev}).

Unlike a superobserver, an outside observer, where only the exterior
metric is 
relevant, sees branes and anti-branes.  Moreover, the momentum $p_{l_5}$ 
corresponding to $Q_5$ is no longer left moving if the
D-fivebrane intersects as an anti-brane. Instead it is right moving, {\it
i.e.}
$u_5\equiv v$. From the outside observer's point of view, the classical 
six-dimensional black string solution, with a new
interpretation of the momentum operator in (\ref{pl15},\ref{plmodes}),
{\it
i.e.} with the D-fivebrane as an anti-brane and with right as well as left
moving momentum, represents the non-extremal case.\footnote{Although we
have 
only considered left and right moving momentum modes and argued that 
(\ref{pl15}) corresponds to the non-extremal case, it seems that the BPS 
supersymmetric properties of the configuration are also different for a 
superobserver and an outside observer.  We leave the detailed study of BPS 
bound state configurations for these two observers for future research.}
The 
outside observer therefore sees the extremal horizon as if it were a 
non-extremal one. The corresponding Bekenstein-Hawking entropy is 
\cite{hor,hor2}
\be
S_{n-e} = 2\pi \sqrt{Q_1Q_5}\left(\sqrt{N_1}+\sqrt{N_5}\right)
\label{nonext}\ee
where $N_{1,5}$ is the number of left,right moving modes. In the absence
of
right moving modes, (\ref{nonext}) becomes the entropy (\ref{extentropy})
of
the extremal black string. If the number of right movers is small, then
the
perturbation away from a purely left-moving extremal background is small.
When
a small number $\delta N_5$ of right moving oscillations is added, the
change
in entropy from the extremal case is
\be
\frac{\Delta{S}}{S_e} = \sqrt{\frac{\delta N_5}{N}}
\ee
where $N \equiv N_1 - N_5$. In terms of the mass change $\delta M$ due to
the
transition to near-extremality, we have \cite{mal}
\be
\frac{\Delta{S}}{S_e}\sim\sqrt{\frac{\delta M}{M_e}}
\label{entropynear}\ee
where $M_e$ is the mass of the extremal black string.

The mass gap between the near-extremal and the extremal black string is
\cite{mal}
\be
 \frac{(\Delta p)^2}{M_e} \equiv \delta M \sim \frac{1}{Q_1 Q_5}
\ee
The entropy difference (\ref{entropynear}) thus coincides with the
deviation
(\ref{entropydev}) caused by the quantum fluctuations of the momentum of
the
stationary black string.  As pointed out before, the decomposition
(\ref{pl15})
of the momentum operator into two independent components does not change
the
result (\ref{dev}). It is the area from the classical black string
solution
that is modified so as to include a right moving momentum, yielding an
identical entropy deviation in both the classical and D-brane pictures. 
Therefore the discrepancy in the black hole/D-brane correspondence can be 
explained from the fact that, unlike a superobserver, the outside observer
does not have access to the interior of black holes.  In other words, the 
discrepancy results from the projection of the augmented electric-magnetic 
symmetric Hilbert space onto physically observable states.

It would be interesting to see how the above interpretation of the D-brane
Hilbert space based on anomalous couplings and electric-magnetic symmetry
can 
be used to describe other issues in black hole physics. For instance, in
\cite{myers} it is argued that pure quantum states do not form black
holes,
which suggests the need for quantum decoherence in the black hole
description.
What we have shown above suggests that there are interactions between the
interior and exterior regions of a black hole at the horizon. Therefore it
seems plausible to interpret the interior of a black hole as the
environment or
external source in quantum decoherence. The quantum corrections, due to
the
external sources, to the entropy of four-dimensional Einstein-Yang-Mills
black
holes are described in \cite{mavromatos}. Furthermore, the decomposition
(\ref{hilbertsp}) of the Hilbert space of the quantum field theory into
two
independent components is the natural setting for the noncommutative
geometry
description of duality symmetries \cite{lizzi}. The approach above may
therefore also be relevant to the noncommutative short-distance structure,
which is inherent in D-brane field theory \cite{wittenD}, of black hole
spacetimes.

\end{document}